\documentclass[
aps,
nofootinbib,
superscriptaddress,
tightenlines,
notitlepage,
twocolumn,
showpacs,
floatfix,
]{revtex4-1}
\usepackage{amssymb,amsmath,bm,tensor,braket}
\usepackage[colorlinks]{hyperref}

\usepackage{graphicx}
\usepackage{dcolumn}
\usepackage{bm}
\usepackage{xcolor}

\newcommand{\PlanckMass}{M_{\rm Pl}}

\newcommand{\Tabriz}{\affiliation{Faculty of Physics, University of Tabriz,
Tabriz 51666-16471, Iran}}
\newcommand{\KIAA}{\affiliation{Kavli Institute for Astronomy and
Astrophysics, Peking University, Beijing 100871, China}}
\newcommand{\NAOC}{\affiliation{National Astronomical Observatories,
Chinese Academy of Sciences, Beijing 100012, China}}

\begin{document}

\preprint{APS/123-QED}

\title{Cosmology and Perturbations in Tachyonic Massive Gravity}

\author{Amin Rezaei Akbarieh}\email{am.rezaei@tabrizu.ac.ir}\Tabriz
\author{Sobhan Kazempour}\email{s.kazempour@tabrizu.ac.ir}\Tabriz
\author{Lijing Shao}\email{lshao@pku.edu.cn}\KIAA\NAOC


\date{\today}

\begin{abstract}
As massive gravity and its extensions offer physically well-defined gravitational theories with a nonzero graviton mass, we present a new extension of the de Rham-Gabadadze-Tolley (dRGT) massive gravity, which is tachyonic massive gravity theory. We firstly introduce the new extension of the dRGT massive gravity, constructed by adding a tachyonic term. We then find the cosmological background equations, and present the analysis of self-accelerating solutions. We examine the tensor perturbations to calculate the dispersion relation of gravitational waves (GWs). In a special case, we consider a constant tachyon potential for the tachyon field $\sigma$, $V(\sigma)={2\omega}/{\PlanckMass^{2}}$, and calculate the equations of motion and the self-accelerating solutions. Finally, we investigate the background perturbations, which include tensor, vector, and scalar perturbations in this case. We calculate the dispersion relation of GWs in the Friedmann-Lema\^itre-Robertson-Walker (FLRW) cosmology in a tachyonic massive gravity theory. These analyses provide potential inputs to future applications in cosmology and GW propagations.
\end{abstract}


\maketitle


\section{Introduction}\label{intro}

One of the most important puzzles in cosmology is the origin of late-time acceleration of the Universe \cite{SupernovaSearchTeam:1998fmf,SupernovaCosmologyProject:1998vns}. Furthermore, there are enough motivations to modify general relativity in order to explain the late-time acceleration of the Universe without a dark energy component \cite{Clifton:2011jh,Carroll:2004de}. In fact, for finding alternative theories to general relativity by maintaining Lorentz invariance, one possible way is considering gravity as a representation of a higher spin \cite{Vasiliev:1995dn}. It is believed that the massive gravity theory is a valuable modification of general relativity by considering the graviton with a non-zero mass \cite{deRham:2014zqa}. While general relativity can be considered as a unique theory of a massless spin-2 particle, in the massive gravity theory a spin-2 massive graviton responds to the propagation of gravity. Also, the speed of gravitational wave (GW) propagation should be determined by the mass of graviton and, this issue can give us an opportunity to constrain the modified gravity theories from recent GW observations~\cite{LIGOScientific:2017zic}. 

Considering a mass to the graviton could be a better scenario for explaining the late-time acceleration of the Universe in comparison with the cosmological constant \cite{deRham:2014zqa,deRham:2010kj}. Fierz and Pauli introduced a massive spin-2 field theory which includes a specific combination of the mass terms to have five physical degrees of freedom \cite{Fierz:1939ix}. However, there is a discontinuity such that the theory in the massless limit does not reduce to the massless theory, i.e., the van Dam-Veltman-Zakharov (vDVZ) discontinuity \cite{Zakharov:1970cc,vanDam:1970vg}. Later, this discontinuity was solved by considering the nonlinear completions of the Fierz-Pauli mass term \cite{Vainshtein:1972sx}. According to studies by Boulware and Daser \cite{Boulware:1972yco}, a sixth ghost degree of freedom appears at the nonlinear level, which is called the Boulware-Daser ghost. Therefore, the nonlinear massive gravity was considered as an unstable theory. Hamed-Arkani et~al.\ and Creminelli et~al.\ obtained the same conclusion \cite{ArkaniHamed:2002sp,Creminelli:2005qk}. However, in 2010, de Rham, Gabadadze and Tolley (dRGT) revisited the analysis and proposed a ghost-free massive gravity theory, which is called the dRGT theory \cite{deRham:2010ik,deRham:2010kj}.  Nevertheless, while this theory admits only an open Friedmann-Lema\^itre-Robertson-Walker (FLRW) solution, the scalar and vector perturbations around the background would vanish. In other words, this problem has something to do with a strong coupling problem and a nonlinear ghost instability \cite{Gumrukcuoglu:2011zh,DeFelice:2012mx}.

Thus, new extensions of the nonlinear massive gravity theory are imperative in finding a stable cosmological solution which is another motivation of our study. One way is to introduce new fields to the dRGT theory, for instance, adding the quasi-dilaton term, mass-varying massive gravity, and bigravity \cite{DAmico:2012hia,Huang:2012pe,Hassan:2011zd,Akbarieh:2021vhv}. In this paper, we add a tachyonic term to the dRGT massive gravity theory as a new extension of the nonlinear massive gravity theory. We try to show the theory which can explain the accelerated expansion of the Universe in FLRW cosmology and would be free of strong coupling problem and a nonlinear ghost instability in perturbations.

Note that a tachyonic scalar field can explain the expansion in the very early universe and the late-time acceleration of the Universe, and its equation of state parameter for dark energy is between $-1$ and 0 \cite{Padmanabhan:2002cp,Gibbons:2002md,Frolov:2002rr,Sen:2003mv}. Moreover, we know that a tachyon field arises in the context of string theory, i.e., tachyon in the Dirac-Born-Infeld action \cite{Sen:2002nu,Sen:2002in,Sen:2002an}. The tachyon has widely been investigated throughout the last few years in cosmological applications and inflaton at high energy \cite{Mazumdar:2001mm,Fairbairn:2002yp,Feinstein:2002aj,Sen:1999xm,Sen:1998sm}.  Meanwhile, there have been recent studies that are related to the tachyonic field and tachyon inflation \cite{Motavalli:2016gid,Gao:2018tdb,Junghans:2016abx,RezaeiAkbarieh:2018ijw,Percacci:2020ddy}.

It is worth mentioning that there has been a trend towards extending the dRGT massive gravity theory, which is free of the Boulware-Deser ghost. These actions admit self-accelerating solutions in which the Universe is of de Sitter type without the cosmological constant \cite{deRham:2010tw,Koyama:2011xz,Nieuwenhuizen:2011sq,DAmico:2011eto}. In addition, it is important to note that the Hubble scale of these self-accelerating solutions is of order of the mass of graviton. According to this fact, having a light graviton is technically natural \cite{Arkani-Hamed:2002bjr,deRham:2012ew}. These solutions are very interesting to be considered for the late-time acceleration of the Universe. Besides, it is insightful to ask how the perturbations around any non-trivial solution behave, and we are interested in finding the new effects in the propagation of the associated degrees of freedom. Perturbation theory for the self-accelerating solutions has been investigated in Refs.~\cite{Gumrukcuoglu:2011zh,DAmico:2012qbv,Wyman:2012iw,EmirGumrukcuoglu:2014uog,Heisenberg:2015voa,Kenna-Allison:2020egn,Kenna-Allison:2019tbu}.

In this paper, we introduce a new extension of the dRGT massive gravity theory, which is obtained by adding a tachyonic field. We analyse the cosmology and perturbations of this new extension of massive gravity theory. The paper is organized as follows. In Sec. \ref{Tach-1}, we present the tachyonic massive gravity theory, and we obtain background equations. In the following step, we derive the self-accelerating solutions and present the tensor perturbation analysis for determining the dispersion relation of GWs. In Sec. \ref{Sp-C}, we examine the background equations for a particular case and demonstrate the self-accelerating solutions. Also, we analyse the cosmological perturbations to show the tensor, vector, and scalar perturbations elaborately. Finally, in Sec. \ref{Con}, some discussion and conclusion are given.

\section{Tachyonic massive gravity theory} \label{Tach-1}

In this section, we introduce a tachyonic massive gravity and its analysis. First, we review the tachyonic massive gravity theory, and we show the evolution of a cosmological background. Then, we derive the self-accelerating solutions. Subsequently, tensor perturbations in this theory are demonstrated.

We use $g_{\mu\nu}$ to represent the background FLRW metric, which is defined as
\begin{eqnarray}
g_{\mu\nu}dx^{\mu} dx^{\nu}= -N^{2}dt^{2}+a^{2}\delta_{ij}dx^{i}dx^{j},
\end{eqnarray}
where, the scale factor is represented by $a$, and $\dot{a}$ is the derivative
with respect to time; $N$ is the lapse function of the dynamical metric, and it
is similar to a gauge function. The lapse function relates the coordinate time
$dt$ and the proper time $d\tau$ via $d\tau =N dt$
\cite{Scheel:1994yr,Christodoulakis:2013xha}. 

We write the action for the tachyonic massive gravity theory in the following
form
\begin{eqnarray}\label{3}
S= \frac{\PlanckMass^{2}}{2}\int d^{4}x\sqrt{-g}\bigg[R-2\Lambda +2m_{g}^{2}\mathcal{L}_{g}\nonumber\\-V(\sigma)\sqrt{1+g_{\mu\nu}\partial^{\mu}\sigma\partial^{\nu}\sigma}\bigg],
\end{eqnarray}
where $\PlanckMass$ is the Planck mass, $R$ is the Ricci scalar, $\Lambda$ is
the cosmological constant, and $V(\sigma)$ is tachyonic potential with the tachyonic field $\sigma (t)$. It is clear that the potential $\mathcal{L}_{g}$ is
the part of Lagrangian which provides the mass to the graviton and can be
written as 
\begin{equation}
   \mathcal{L}_{g}=\mathcal{L}_{2}+\alpha_{3}\mathcal{L}_{3}+\alpha_{4}\mathcal{L}_{4} ,
\end{equation}
where $\alpha_{3}$ and $\alpha_{4}$ are dimensionless free parameters of the
theory. In the equation, $\mathcal{L}_{i}$ $(i=2,3,4)$ are given by
\begin{eqnarray}
\mathcal{L}_{2}&=&\frac{1}{2}([\mathcal{K}]^{2}-[\mathcal{K}^{2}]),
 \nonumber\\
 \mathcal{L}_{3}&=&\frac{1}{6}([\mathcal{K}]^{3}-3[\mathcal{K}][\mathcal{K}^{2}]+2[\mathcal{K}^{3}]),
 \nonumber\\
 \mathcal{L}_{4}&=&\frac{1}{24}([\mathcal{K}]^{4}-6[\mathcal{K}]^{2}[\mathcal{K}^{2}]+8[\mathcal{K}][\mathcal{K}^{3}]+3[\mathcal{K}^{2}]^2  \nonumber\\&& -6[\mathcal{K}^{4}]).
\end{eqnarray}
The square brackets donate a trace of $\mathcal{K}^{\mu}_{\nu}$. The above terms
are similar to that in the dRGT theory \cite{deRham:2010kj}. We have defined the
building block tensor $\mathcal{K}^{\mu}_{\nu}$ as
\begin{eqnarray}
\mathcal{K}^{\mu}_{\nu}=\delta^{\mu}_{\nu}-e^{\sigma/\PlanckMass}\big(\sqrt{g^{-1}\bar{g}}\big)^{\mu}_{\nu}.
\end{eqnarray}
We use $\bar{g}_{\mu\nu}$ as a non-dynamical fiducial metric and we use
Minkowski metric as the fiducial metric,
\begin{eqnarray}
\bar{g}_{\mu\nu}=-\dot{f}(t)^{2}dt^{2}+\delta_{ij}dx^{i}dx^{j},
\end{eqnarray}
where $f(t)$ is the Stueckelberg scalar.

After considering the equations above, the total Lagrangian of the tachyonic
massive gravity in a FLRW universe is
\begin{eqnarray}
\mathcal{L}&=&\PlanckMass^{2}\bigg[-3\frac{a\dot{a}^{2}}{N}-\Lambda{a}^{3}N\bigg]
+m_{g}^{2}\PlanckMass^{2}\bigg\{N a^{3}(X-1)\nonumber\\
&&\times\big[3(X-2)-(X-4)(X-1)\alpha_{3}-(X-1)^{2}\alpha_{4}\big]\nonumber\\
&&+\dot{f}(t)a^{4}X(X-1)\Big[3-3(X-1)\alpha_{3}+(X-1)^{2}\alpha_{4}\Big]\bigg\}\nonumber\\
&&-\frac{1}{2}a^{3}\PlanckMass^{2}V(\sigma)N\sqrt{1-\frac{\dot{\sigma}^{2}}{N^{2}}},
\end{eqnarray}
where $X\equiv \frac{e^{\sigma/\PlanckMass}}{a}$.

\subsection{Background equations}

Note that we obtain a constraint equation by varying the Lagrangian with respect to $f$,
\begin{eqnarray}\label{8}
m_{g}^{2}\PlanckMass^{2}\frac{d}{dt}\Bigg\{a^{4}X(X-1)\bigg[3+3(1-X)\alpha_{3}\nonumber\\+(1-X)^{2}\alpha_{4}\bigg]\Bigg\}=0.
\end{eqnarray}

In order to calculate the Friedman equation, we take the variation of the
Lagrangian with respect to $N$ and obtain,
\begin{eqnarray}\label{9}
3H^{2}-\Lambda -\frac{1}{2}\frac{V(\sigma)}{\sqrt{1-\frac{\dot{\sigma}^{2}}{N^{2}}}}
+m_g^{2}(X-1)\bigg[3(X-2)\nonumber\\-\alpha_{3}(X-4)(X-1)-\alpha_{4}(X-1)^{2}\bigg]=0,
\end{eqnarray}
where $H\equiv \frac{\dot{a}}{Na}$ and
$\frac{\dot{\sigma}}{N}=\PlanckMass(H+\frac{\dot{X}}{NX})$. 

In the following, the acceleration equation can be obtained by varying with
respect to the scale factor $a$,
\begin{eqnarray}\label{10}
3H^{2}&&-\Lambda+\frac{2\dot{H}}{N}-\frac{V(\sigma)}{2}\sqrt{1-\frac{\dot{\sigma}^{2}}{N^{2}}}
 +m_{g}^{2}\bigg\{6-3(2 +r)X  \nonumber\\
&& +(1+2r)X^{2}-(X-1) \Big[\alpha_{3}\big(4-(2+3r)X+rX^{2}\big) \nonumber\\
&&  +\alpha_{4}(X-1)(rX-1)\Big]\bigg\}=0.
\end{eqnarray}
In the following, we consider $N=1$ and, we define $r=\frac{a\dot{f}}{N}$. 

Finally, varying the Lagrangian with respect to  $\sigma(t)$ gives 
\begin{eqnarray}\label{11}
&& \ddot{\sigma}+\bigg(3H\dot{\sigma}+\frac{V^{'}(\sigma)}{V(\sigma)}\bigg)\big(1-\dot{\sigma}^{2}\big)
-\frac{6 m_{g}^{2}(1-\dot{\sigma}^{2})^{3/2}}{\PlanckMass V(\sigma)}  \times \nonumber\\
&& X\Bigg\{\Big[2X-3+r(2X-1)\Big]+(X-1)\bigg[\alpha_{3} \Big( r(1-3X) \nonumber\\
&& +3-X \Big)-\frac{1}{3}\alpha_{4}(X-1)\Big(3+r(1-4X)\Big)\bigg]\Bigg\} =0.
\end{eqnarray}
It should be pointed out that the Stuckelberg field $f$ introduces time
re-parametrization invariance. As a result, there is a Bianchi identity which
relates the four equations of motion, 
\begin{eqnarray}
\frac{\delta S}{\delta \sigma}\dot{\sigma}+\frac{\delta S}{\delta f}\dot{f}-N\frac{d}{dt}\frac{\delta S}{\delta N}+\dot{a}\frac{\delta S}{\delta a}=0.
\end{eqnarray}
Therefore, one equation is redundant and can be eliminated.

We should take this into consideration, and if we consider $V(\sigma)=0$ in the particular conditions, the theory, all above background equations, and the total Lagrangian, reduce to those in Ref.~\cite{Gumrukcuoglu:2011zh}

\subsection{Self-accelerating regime}\label{Fi-Se}

In this subsection, the self-accelerating solutions are discussed. It should be mentioned that using the asymptotic background equations of motion in the self-accelerating regime, which was considered in Ref.~\cite{deRham:2010ik}, we will simplify the expression of $M^{2}_{\rm GW}$
in the next section. 

By varying the action with respect to $f$ and considering the
Eq.~(\ref{8}), we  find the constraint equation $J = \frac{1}{a^{4}}\times {\rm
constant}$, where 
\begin{eqnarray}
J=X(X-1)\big(3+3(1-X)\alpha_{3}+(1-X)^{2}\alpha_{4}\big).
\end{eqnarray}
In an expanding Universe, the right-hand side of this equation decay as
$a^{-4}$, hence after a sufficiently long time, $X$ saturates to a constant
value $X_{\rm SA}$, corresponding to $J=0$. These constant solutions of $X$ lead
to the effective energy density and act like a cosmological constant. As
discussed in Ref.~\cite{DAmico:2012hia}, there are four such solutions for which
$X$ is constant. We should pay attention that $X_{\rm SA}^{0} = 0$  implies
$\sigma \rightarrow \infty$, and as in Ref.~\cite{DAmico:2012hia}, we disregard
this solution in order to avoid strong coupling. Three other solutions  can be
obtained for the state of the accelerated Universe. One of these solutions is
$X^{1}_{\rm SA}=1$, and the other two are

\begin{equation}
X_{\rm SA}^{\pm}=\frac{(\alpha_{3}+2\alpha_{4})\pm \sqrt{\alpha_{3}^{2}-12\alpha_{4}}}{2\alpha_{4}}.
\end{equation}

As one can see, if we choose $X^{1}_{\rm SA}=1$ and substitute it in
Eq.~(\ref{9}), the mass term $m_{g}$  becomes zero. So we are not interested in
this solution. In other words, in this case, the constant term of the effective
cosmology becomes zero, and the background will be de Sitter mode. However, in
the presence of matter fields and without a bare cosmological constant, this
solution asymptotically approaches a Minkowski background and is unstable
\cite{DAmico:2012hia}. So we will consider $X_{\rm SA}^{\pm}$ solutions.
Substituting these solutions in Eq.~(\ref{9}), we obtain the generalized
Friedman equation,
\begin{equation}\label{15}
3H^{2}-\frac{V(\sigma)}{2\sqrt{1-(\PlanckMass H)^{2}}}= \Lambda+\Lambda_{\rm SA}^{\pm},
\end{equation}
where
\begin{eqnarray}\label{16}
\Lambda_{\rm SA}^{\pm} \equiv && \frac{1}{2\alpha_{4}^{3}}\bigg[\alpha_{3}^{4}\pm\alpha_{3}^{3}\Big(\sqrt{\alpha_{3}^{2}-12\alpha_{4}}\pm 4\alpha_{4}\Big)   \pm \alpha_{3}^{2}\alpha_{4} \times \nonumber\\
&& \Big(\pm 3
+2\sqrt{\alpha_{3}^{2}-12\alpha_{4}}\Big)+24\alpha_{3}\alpha_{4}^{2}-18\alpha_{4}^{2}\bigg].
\end{eqnarray}

In Eq.~(\ref{15}), if we consider the potential as a constant function such as
$V(\sigma)=\frac{\omega}{\PlanckMass^{2}}$, we have
\begin{eqnarray}\label{17}
3H^{2} =\Lambda_{\rm eff}^{\pm},
\end{eqnarray}
where
\begin{eqnarray}
\Lambda_{\rm eff}^{\pm}=\frac{1}{6\PlanckMass^{2}}\bigg[6+4 \big(\Lambda &&+\Lambda_{\rm SA}^{\pm} \big)+\Xi^{\pm 2}\nonumber\\&&+\frac{4\big(-3+\Lambda+\Lambda_{\rm SA}^{\pm}\big)^{2}}{\Xi^{\pm}}\bigg],
\end{eqnarray}
with
\begin{eqnarray}
\Xi^{\pm}=&&\bigg[-8\big(-3+\Lambda +\Lambda_{\rm SA}^{\pm}\big)^{3}-81\omega^{2}\nonumber\\&&+9\sqrt{16\big(-3+\Lambda+\Lambda_{SA}^{\pm}\big)^{3}\omega^{2}+81\omega^{4}}\bigg]^{1/3}.
\end{eqnarray}
From Eq.~(\ref{11}), $r_{\rm SA}^{\pm}$ is obtained as
\begin{eqnarray}\label{Self-Acc}
\hspace{-0.68cm} r_{\rm SA}^{\pm}  = 1+  \frac{\PlanckMass \bigg[3 H^{2}V(\sigma)\PlanckMass +V^{'}(\sigma)\bigg]}{6m_{g}^{2}X_{\rm SA}^{\pm ~2}\sqrt{1-H^{2}\PlanckMass^{2}}\bigg[\alpha_{3}X_{\rm SA}^{\pm}-\alpha_{3}-2\bigg]}.
\end{eqnarray}
The above equation can interpret the self-accelerating universe according to new parameters of the theory with considering that there is not any strong coupling. \\
Here, we demonstrate that this theory possesses self-accelerating solutions with an effective cosmological constant is given by $\Lambda_{\rm eff}^{\pm}$.

\subsection{Tensor perturbations}

The study of cosmological perturbations is a very important pillar for modern
cosmology. By examining perturbations, we can find the relationship between
early cosmic models and the present ones.  In this subsection, we consider
tensor perturbations in the tachyonic massive gravity. 

We introduce tensor perturbations in the metric $g_{\mu\nu}$ as $\delta
g_{ij}=a^{2}h_{ij}$, where the perturbations $h_{ij}$ are defined to be
transverse and traceless, i.e. $\partial^{i}h_{ij} = 0$ and $g^{ij}h_{ij} = 0$ \cite{Kahniashvili:2014wua,DeFelice:2013tsa}.
Here, we lower and raise spatial indices on $h_{ij}$ with $\delta_{ij}$ and
$\delta^{ij}$. While we should always raise and lower indices with the metric,
using the Kronecker delta does not affect the second-order action. So the
transverse and traceless conditions can be rewritten as
$\delta^{ik}\partial_{k}h_{ij}=0$ and $\delta^{ij}h_{ij} = 0$. Also, it is
interesting to note that the perturbed metric $\delta g_{\mu\nu}$ is always
defined to be first order. However, the inverse metric $g^{\mu\nu}$ can be of
higher order.

Applying tensor perturbations to the gravitational part of the action, 
\begin{eqnarray}
S_{\rm
gravity}=\frac{M_{\rm Pl}^2}{2}\int d^{4}x\sqrt{-g}\big(R-2\Lambda\big),
\end{eqnarray}
we have
\begin{eqnarray}
S^{(2)}_{\rm gravity}=&&\frac{\PlanckMass^{2}}{8}\int d^{3}k \, dt \, a^{3}N\bigg[\frac{\dot{h}_{ij}\dot{h}^{ij}}{N^{2}}\nonumber\\
&&-\Big(\frac{k^{2}}{a^{2}}+4\frac{\dot{H}}{N}+6H^{2}-2\Lambda\Big)h_{ij}h^{ij}\bigg].
\end{eqnarray}
We have used the Fourier transform convention $h_{ij}(\vec{x},t)=\int
\frac{d^{3}k}{(2\pi)^{3/2}}h_{ij}(\vec{k}, t)e^{-i\vec{k} \cdot \vec{x}}$.  For
the tachyonic part of the action,
\begin{eqnarray}
S_{\rm tachyon}=-\frac{\PlanckMass^{2}}{2}\int d^{4}x\sqrt{-g}V(\sigma)\sqrt{1+g_{\mu\nu}\partial^{\mu} \sigma\partial^{\nu}\sigma},\nonumber\\
\end{eqnarray}
we obtain
\begin{eqnarray}
&&S^{(2)}_{\rm tachyon}=\nonumber\\ 
&&\frac{\PlanckMass^{2}}{8}\int d^{3}k  \, dt \, a^{3}N\bigg[V(\sigma)\sqrt{1-\frac{\dot{\sigma}^{2}}{N^{2}}}\bigg]h_{ij}h^{ij}.
\end{eqnarray}
For the massive gravity part, we find
\begin{eqnarray}
S^{(2)}_{\rm massive}=&&\frac{\PlanckMass^{2}}{8}\int d^{3}k \, dt \, a^{3}N m_{g}^{2} \bigg\lbrace(\alpha_{3}+\alpha_{4})rX^{3}\nonumber\\
&&-(1+2\alpha_{3}+\alpha_{4})(1+3r)X^{2}+(3+3\alpha_{3}+\alpha_{4})\nonumber\\
&& \times (3+2r)X-2(6+4\alpha_{3}+\alpha_{4})\bigg\rbrace h_{ij}h^{ij},
\end{eqnarray}
where $r=\frac{a\dot{f}}{N}$. Then the second-order action for tensor perturbations is obtained as
\begin{eqnarray}
S^{(2)}=&&\frac{\PlanckMass^{2}}{8}\int d^{3}k \, dt \, a^{3}N \nonumber\\
&& \times \Bigg[\frac{\dot{h}^{ij}\dot{h}_{ij}}{N^{2}}  -\Big(\frac{k^{2}}{a^{2}}+M^{2}_{\rm GW}\Big)h_{ij}h^{ij}\Bigg].
\end{eqnarray}
Here we consider $N=1$, therefore, the dispersion relation of GWs is given by,
\begin{eqnarray}\label{MG1}
\hspace{-0.5cm} M^{2}_{\rm GW}&&=4\dot{H}+6H^{2}-2\Lambda -V(\sigma)\sqrt{1-\dot{\sigma}^{2}}\nonumber\\
&& -m^{2}_{g}\Bigg[(\alpha_{3} +\alpha_{4})rX^{3}-(1+2\alpha_{3}+\alpha_{4})(1+3r)X^{2}\nonumber\\
&& +(3 +3\alpha_{3}+\alpha_{4})(3+2r)X-2(6+4\alpha_{3}+\alpha_{4})\Bigg].
\end{eqnarray}
Note that if we consider $V(\sigma)=0$ in the particular conditions, we find the same results for the dispersion relation of GWs as in Ref.~\cite{Gumrukcuoglu:2011zh,DeFelice:2013tsa,Gumrukcuoglu:2013nza,DAmico:2013saf}.

\section{A special case}\label{Sp-C}

It is worth pointing out that the tachyon can have a role as a part of the source of dark energy according to the form of the tachyon potential. There are several tachyon potentials that are considered for the open string theory, the non-BPS D-brane in the super-string, the bosonic string theory, and another type of tachyon potential \cite{Padmanabhan:2002cp,Garousi:2004uf,Garousi:2004hy,Chingangbam:2004xe,Bagla:2002yn,Abramo:2003cp,Aguirregabiria:2004xd,Copeland:2004hq,Kutasov:2003er}.
Here, we consider a special case in detail where the tachyon potential is a
constant, $V(\sigma)=\frac{2\omega}{\PlanckMass^{2}}$.

In the following stages, we examine the evolution of a cosmological background
for the special case and present the self-accelerating solutions as well.
Meanwhile, we show cosmological perturbation analyses which consist of tensor,
vector, and scalar perturbations.

\subsection{Background equations}

We start by finding the equations of motion. Varying the action, the energy
density and pressure are obtained as
\begin{eqnarray}
\rho =\frac{\omega}{\sqrt{1-\dot{\sigma}^{2}}}, \qquad P =-\omega \sqrt{1-\dot{\sigma}^{2}}.
\end{eqnarray}
Then, the sound speed is
\begin{eqnarray}
c_{S}^{2} =\dot{\sigma}^{2}-1.
\end{eqnarray}
In addition, we fix the cosmological time using $N=1$. So, the equation of
motion with respect to $N$ is
\begin{eqnarray}\label{EoM-N-Se}
3H^{2}=\Lambda +\frac{\rho}{\PlanckMass^{2}}+m_{g}^{2}L,
\end{eqnarray}
where
\begin{eqnarray}
L\equiv -(6+4\alpha_{3}+\alpha_{4})+3(3+3\alpha_{3}+\alpha_{4})X\nonumber\\-3(1+2\alpha_{3}+\alpha_{4})X^{2}+(\alpha_{3}+\alpha_{4})X^{3}.
\end{eqnarray}
By varying with respect to $a$, we obtain
\begin{eqnarray}\label{EoM-A.Se}
2\dot{H}=m_{g}^{2}Q(r-1)X-\frac{\rho +P}{\PlanckMass^{2}},
\end{eqnarray}
where
\begin{eqnarray}
Q\equiv 3+3\alpha_{3}+\alpha_{4}-2(1+2\alpha_{3}+\alpha_{4})X\nonumber\\+(\alpha_{3}+\alpha_{4})X^{2}.
\end{eqnarray}
Also, by varying the Lagrangian with respect to $\sigma(t)$, we obtain
\begin{eqnarray}
\ddot{\sigma}  && +3H\dot{\sigma}(1-\dot{\sigma}^{2}) \nonumber\\ &&+\frac{\PlanckMass}{\omega}m_{g}^{2}(1-\dot{\sigma}^{2})^{3/2}X\big[3Q
+rK\big]=0,
\end{eqnarray}
where
\begin{eqnarray}
K\equiv 3+3\alpha_{3}+\alpha_{4}-6(1+2\alpha_{3}+\alpha_{4})X\nonumber\\+9(\alpha_{3}+\alpha_{4})X^{2}-4\alpha_{4}X^{3}.
\end{eqnarray}

\subsection{Self-accelerating regime}

Here, we should point out that the result of this special case is similar to the
previous case in Sec.~\ref{Fi-Se}.  For an accelerated expansion of the
Universe, we have $J=0$. which leads to 
\begin{eqnarray}
X_{\rm SA}^{\pm}=\frac{1}{2\alpha_{4}}\Big(3\alpha_{3}+2\alpha_{4}\pm \sqrt{9\alpha_{3}^{2}-12\alpha_{4}}\Big).
\end{eqnarray}
In this case, we have
\begin{eqnarray}
L_{\rm SA}^{\pm}=&&\frac{3}{2\alpha_{4}^{3}}\bigg\lbrace 9\alpha_{3}^{4} -18\alpha_{3}^{2}\alpha_{4}+6\alpha_{4}^{2}\nonumber\\&&\pm(3\alpha_{3}^{2}-4\alpha_{3})\sqrt{9\alpha_{3}^{2}-12\alpha_{4}}\bigg\rbrace .
\end{eqnarray}
According to Eq.~(\ref{EoM-N-Se}),  $L_{\rm SA}^{\pm}$ is similar to an
effective cosmological constant. In other words, it can be considered as an
effective energy density, $m_{g}^{2}\PlanckMass^{2}L_{\rm SA}^{\pm}$, which
arises from the mass term.

Furthermore, in the  accelerated expanding regime, we have
\begin{eqnarray}
\hspace{-0.5cm}
Q_{\rm SA}^{\pm}=&&3+3\alpha_{3}+\alpha_{4}\mp\frac{1}{\alpha_{4}}(1+2\alpha_{3}+\alpha_{4}) \nonumber\\
&& \times \bigg[\pm(3\alpha_{3} +2\alpha_{4})+\sqrt{9\alpha_{3}^{2}-12\alpha_{4}}\bigg]\nonumber\\
&& +\frac{(\alpha_{3}+\alpha_{4})}{4\alpha_{4}^{2}}\bigg(3\alpha_{3} +2\alpha_{4}\pm\sqrt{9\alpha_{3}^{2}-12\alpha_{4}}\bigg)^{2}.
\end{eqnarray}
According to Eq.~(\ref{EoM-A.Se}), it can be concluded that
$m_{g}^{2}\PlanckMass^{2}Q_{\rm SA}^{\pm}(1-r_{\rm SA}^{\pm})X_{\rm SA}^{\pm}$
is related to the sum of the effective density and pressure.  Therefore, we have
\begin{eqnarray}
K_{\rm SA}^{\pm}=3+3\alpha_{3}+\alpha_{4}-6(1+2\alpha_{3}+\alpha_{4})X_{\rm SA}^{\pm} \nonumber\\+9(\alpha_{3}+\alpha_{4})X_{\rm SA}^{\pm ~2}-4\alpha_{4}X_{\rm SA}^{\pm ~3}.
\end{eqnarray}
For simplicity, we use $\tilde{Q}=m_{g}^{2}QX$ and $\tilde{Q}_{\rm
SA}^{\pm}=m_{g}^{2}Q_{\rm SA}^{\pm}X_{\rm SA}^{\pm}$ throughout this paper.
Thus, we have
\begin{eqnarray}\label{Self-r2}
r_{\rm SA}^{\pm}=\frac{1}{\PlanckMass^{2}\bigg(2K_{\rm SA}^{\pm}m_{g}^{2}(1-H^{2}\PlanckMass^{2})^{3/2}X_{\rm SA}^{\pm}+\omega \tilde{Q}_{\rm SA}^{\pm}\bigg)}\nonumber\\\Bigg\lbrace -2\Lambda \omega \PlanckMass^{2}-2\omega L_{\rm SA}^{\pm}m_{g}^{2}\PlanckMass^{2}+2\omega \PlanckMass \rho_{\rm SA}(H^{2}\PlanckMass^{2}-1) \nonumber\\+\omega (P_{\rm SA}+\rho_{\rm SA}) 
+2H^{2}\omega L_{\rm SA}^{\pm}m_{g}^{2}\PlanckMass^{4}+\omega \PlanckMass^{2}\tilde{Q}_{\rm SA}^{\pm}\nonumber\\-6\PlanckMass^{2}(1-H^{2}\PlanckMass^{2})^{3/2}\tilde{Q}_{\rm SA}^{\pm}+2H^{2}\Lambda \omega \PlanckMass^{4}\Bigg\rbrace. \nonumber\\
\end{eqnarray}
As we mentioned at the end of subsection (\ref{Fi-Se}), equation (\ref{Self-r2}) could explain the self-accelerating universe in a special case that is free of strong coupling too.
 
\subsection{Cosmological perturbations}

In this stage, we focus on quadratic perturbations. To find the quadratic
perturbations, the physical metric $g_{\mu\nu}$ should be expanded in terms of
small fluctuations $\delta g_{\mu\nu}$ around a background solution
$g_{\mu\nu}^{(0)}$,
\begin{equation}
g_{\mu\nu}=g_{\mu\nu}^{(0)}+\delta g_{\mu\nu}.
\end{equation}
Also, we should keep  terms up to the quadratic order.  The metric perturbations
can be divided into three parts, namely scalar, vector and tensor perturbations.
They are expressed as 
\begin{eqnarray}
\delta g_{00}=&&-2N^{2} \Phi, \nonumber\\
\delta g_{0i}=&&Na(B_{i}+\partial_{i}B), \nonumber\\
\delta g_{ij}=&&a^{2}\bigg[h_{ij}+\frac{1}{2}(\partial_{i}E_{j}+\partial_{j}E_{i})+2\delta_{ij}\Psi \nonumber\\
&& +\big(\partial_{i}\partial_{j} -\frac{1}{3}\delta_{ij}\partial_{l}\partial^{l}\big)E\bigg].
\end{eqnarray}
Here, notice that the tensor perturbations are transverse,
$\partial^{i}h_{ij}=0$, and traceless, $h_{i}^{~i}=0$. Meanwhile, the vector
perturbations have conditions such that $\partial^{i}B_{i}=0$ and
$\partial^{i}E_{i}=0$ \cite{Kahniashvili:2014wua,DeFelice:2013tsa}. In addition, the perturbation of the scalar field is 
\begin{equation}
\sigma =\sigma^{(0)}+\PlanckMass\delta\sigma.
\end{equation}
As we mentioned before, the spatial indices on perturbations can be raised and
lowered by $\delta^{ij}$ and $\delta_{ij}$. The expanded action can be written
in the Fourier domain with plane waves, via $\vec{\nabla}^{2}\rightarrow
-k^{2}$, $d^{3}x\rightarrow d^{3}k$. It should be clarified that we perform all
analysis in the unitary gauge \cite{Gumrukcuoglu:2013nza,Akbarieh:2021vhv}. Therefore, there are not any problems with the form of gauge-invariant combinations. As before, we consider $N=1$ in the
following calculations.

\subsubsection{Tensor perturbation}

The tensor quadratic action is obtained as
\begin{eqnarray}
S^{(2)}=&&\frac{\PlanckMass^{2}}{8}\int d^{3}k \, dt  \, a^{3} \times \nonumber\\
&&\bigg[\dot{h}^{ij}\dot{h}_{ij} -\Big(\frac{k^{2}}{a^{2}}  +M_{\rm GW}^{2}\Big)h^{ij}h_{ij}\bigg],
\end{eqnarray}
where the dispersion relation of GWs is given by
\begin{eqnarray}\label{MGW2}
\hspace{-0.6cm}
M_{\rm GW}^{2}=&&6H^{2}-2\Lambda +\frac{2\omega}{\PlanckMass^{2}\sqrt{1-H^{2}\PlanckMass^{2}}} \nonumber\\
&& +m_{g}^{2}\bigg\{6 +2\big(r-5\big)(X_{\rm SA}^{\pm})^{2}-3\big(r-2\big)(X_{\rm SA}^{\pm})^{3}\nonumber\\
&& +\alpha_{3} \Big[2 +\big(r-5+4X_{\rm SA}^{\pm}-2rX_{\rm SA}^{\pm}\big)(X_{\rm SA}^{\pm})^{2}\Big]\bigg\}.
\end{eqnarray}
In order to obtain the above equation, we have substituted some of the parameters as has been explained below.
The above equation is merely Eq.~(\ref{MG1}). In other words, in the first stage, we have written it using substitution $V(\sigma)=\frac{2\omega}{\PlanckMass^{2}}$. To achieve that equation, we have used the background acceleration in Eq.~(\ref{EoM-A.Se}) to eliminate terms
with $\ddot{a}$. Moreover, the Friedman equation (\ref{9}) is evaluated on the self-accelerating limit, i.e. $\dot{X}=0$ and $\dot{\sigma}=\PlanckMass H$.
Finally, to eliminate $\alpha_{4}$, we have used
\begin{eqnarray}
X_{\rm SA}^{\pm}=\frac{1}{2\alpha_{4}}\Big(3\alpha_{3}+2\alpha_{4}\pm \sqrt{9\alpha_{3}^{2}-12\alpha_{4}}\Big).
\end{eqnarray}
Note that if  $M_{\rm GW}^{2}>0$, the stability of long-wavelength GWs is guaranteed. However, if $M_{\rm GW}^{2}<0$, it will be unstable. In that case, as the mass is of order of the Hubble scale, the instability should take the age of the Universe to develop.

Equations (\ref{MG1}) and (\ref{MGW2}) are the modified dispersion relations of gravitational waves and can be considered as the main result of these parts. They represent the propagation of gravitational perturbations in the FLRW cosmology in the tachyonic massive gravity. These results can be tested by gravitational-wave observations. They will introduce extra contribution to the phase evolution of gravitational waveform \cite{Will:1997bb,Mirshekari:2011yq}, and to be detected with the accurate matched-filtering techniques in the data analysis. The tests of graviton mass have been done, after the first discovery of gravitational waves in a merging binary black hole \cite{LIGOScientific:2016lio,LIGOScientific:2019fpa,LIGOScientific:2020tif,Shao:2020shv}. According to the latest constraint on the graviton mass which is achieved by the combination of gravitational wave events from the first and second gravitational-wave transient catalogs, we know that mass of graviton is around $m_{g} \leq 1.76 \times 10^{-23} \, \mathrm{eV} / c^{2}$ at 90\% credibility~\cite{LIGOScientific:2020tif}. As the Compton wavelength is still much smaller than the Hubble scale, therefore the relevance to modified cosmology is restricted at present. We hope that testing the mass of graviton with gravitational observations at different wavelengths, notably with future space based gravitational-wave detectors which are more sensitive to the graviton mass \cite{Will:1997bb}, can help to find the exact value of the mass of graviton.

\subsubsection{Vector perturbations}

Now we consider the vector perturbations,
\begin{eqnarray}\label{Bi}
B_{i}=\frac{a(1+r)k^{2}}{2\bigg[k^{2}(r+1)+2a^{2}\tilde{Q}\bigg]}\dot{E}_{i}.
\end{eqnarray}
In the above equation, $B_{i}$ is a non-dynamical field and it is calculated
using the equation of motion. By inserting it back into the action, one finds a single
propagating vector
\begin{eqnarray}
\hspace{-0.5cm} S_{\rm vector}^{(2)}=\frac{\PlanckMass^{2}}{8}\int d^{3}k \, dt \, a^{3} 
\bigg(\tau |\dot{E}_{i}|^{2} -\frac{k^{2}}{2}M_{\rm GW}^{2}|E_{i}|^{2}\bigg),
\end{eqnarray}
where
\begin{eqnarray}
\tau =\bigg(\frac{2}{k^{2}}+\frac{1+r}{a^{2}\tilde{Q}}\bigg)^{-1}.
\end{eqnarray}
To avoid ghost and instability in sub-horizon scales we should have
\begin{eqnarray}
\tau \Big|_{k\gg a H}=\frac{a^{2}\tilde{Q}}{1+r}>0.
\end{eqnarray}
The sound speed for the vector modes is
\begin{eqnarray}
c_{V}^{2}=\frac{a^{2}M_{\rm GW}^{2}}{2\tau}\bigg|_{k\gg aH}=\frac{(1+r)M_{\rm GW}^{2}}{2\tilde{Q}}.
\end{eqnarray}
It is understood that, to avoid gradient instability, we need to have $c_{V}^{2}>0$.
Therefore the stability for vector modes is ensured if $c_{V}^{2}>0$. This condition impose $M_{\rm GW}^{2}>0$. So, it constrains the other parameters of the theory.

\subsubsection{Scalar perturbations}

In the scalar perturbations, there are five degrees of freedom, which we denote
as $\Phi$, $B$, $\Psi$, $E$, and $\delta \sigma$. Also, it should be noted that,
the time derivatives of perturbations of lapse $\Phi$ and shift $B$ do not
appear in action. Using the equation of motion, which is related to $B$, one
finds
\begin{eqnarray}
B=&&\frac{(1+r)}{3\PlanckMass a\dot{\sigma}\tilde{Q}}\bigg\lbrace 3\delta\sigma(\rho +P) \nonumber\\
&&+\PlanckMass \dot{\sigma}\big[k^{2}\dot{E} +6(\dot{\Psi}-H\Phi)\big]\bigg\rbrace .
\end{eqnarray}
To calculate the above equation, all perturbations have been expanded with
respect to scalar harmonics. Here, using the lapse perturbation equation, we
can find the auxiliary field $\Phi$, as
\begin{eqnarray}
\Phi &&=\frac{\PlanckMass^{2}c_{s}^{2}a^{2}\tilde{Q}}{2\PlanckMass^{2}c_{s}^{2}H^{2}\bigg[2k^{2}(r+1)+3a^{2}\tilde{Q}\bigg]-a^{2}\tilde{Q}(\rho +P)} \times  \nonumber\\
&&\bigg\lbrace \frac{2k^{2}H(r+1)}{3a^{2}\tilde{Q}}\bigg[\frac{3(\rho +P)}{\PlanckMass \dot{\sigma}}\delta\sigma+k^{2}\dot{E}+3\Big(2  +\frac{3a^{2}\tilde{Q}}{(r+1)k^{2}}\Big)\dot{\Psi}\bigg]  \nonumber\\
&&+\frac{k^{4}}{3a^{2}}E+\frac{2k^{2}+3a^{2}\tilde{Q}}{a^{2}}\Psi -\frac{(\rho +P)}{\PlanckMass c_{s}^{2}\dot{\sigma}}\delta\dot{\sigma}\bigg\rbrace.
\end{eqnarray}
Now with these solutions and by defining  $Y\equiv (\Psi ,E, \delta\sigma)$, the
action can be written as
\begin{eqnarray}
S_{\rm scalar}^{(2)} && =\frac{\PlanckMass^{2}}{2}\int d^{3}k \, dt \, a^{3}  \times \nonumber\\
 && \hspace{-0.3cm} \bigg(\dot{Y}^{\dagger}\Sigma \dot{Y}+\frac{1}{2}\dot{Y}^{\dagger}\Gamma Y+\frac{1}{2}Y^{\dagger}\Gamma^{T}\dot{Y}-Y^{\dagger}MY\bigg),
\end{eqnarray}
where $\Sigma$, $\Gamma$ and $M$ represent the kinetic energy matrix, the mixing
perturbation matrix, and the mass matrix, respectively. As there is no
Boulware-Deser ghost, i.e. ${\rm det} \Sigma =0$, we can eliminate one of the
non-dynamical degrees of freedom. In fact, it is possible to define a quantity
by a combination of $\Psi$, $E$ and $\delta \sigma$, as, 
\begin{eqnarray}
\varsigma =\Psi +\frac{k^{4}(r+1)}{9a^{2}\tilde{Q}+6(r+1)k^{2}}E -\Big(\frac{H}{\dot{\sigma}}\Big)\delta\sigma .
\end{eqnarray}
As the kinetic energy of the combination of perturbation fields is zero, we can
neglect $\delta\sigma$ in comparison with $\varsigma$. Thus, one can see that
the kinetic energy part of the action can be presented diagonally. In addition,
if we consider  $\Psi$, which is an auxiliary component, as a Boulware-Deser
ghost, we can eliminate it too. Therefore, the action can be obtained using two
dynamical fields, encoded in $R=(\varsigma, E)$, as
\begin{eqnarray}\label{56}
S_{\rm scalar}^{(2)}&& = \frac{\PlanckMass^{2}}{2}\int d^{3}k \, dt \times  \nonumber\\
&&\bigg(\dot{R}^{\dagger}\Pi \dot{R}+\frac{1}{2}\dot{R}^{\dagger} \Omega R +\frac{1}{2}R^{\dagger} \Omega^{T}\dot{R}-R^{\dagger}\tilde{M}R\bigg),
\end{eqnarray}
where $\Omega$ represents the mixing and hermitian matrix; $\Pi$ and $\tilde{M}$
are kinetic energy matrix and mass matrix, respectively.

To evaluate Eq. (\ref{56}), we introduce the below discussion. In order to find the ghost-free conditions, we should study the eigenvalue of the kinematic energy matrix $\Pi$ which means that the positivity of this value should be evaluated. This study in the sub-horizon limit (i.e. $ k \rightarrow \infty$) is very crucial. Therefore, we should focus on the components of $\Pi$, in the scalar part, we reach the eigenvalues $\lambda_{1}$ and $\lambda_{2}$ which is given as below
\begin{eqnarray}
\lambda_{1}^{-1}=\frac{\PlanckMass^{4}c_{s}^{2}H^{2}}{2(\rho +P +\PlanckMass^{2}\Lambda)}-\bigg[\frac{8k^{2}(r+1)+12m_{g}^{2}a^{2}LX}{\PlanckMass^{2}m_{g}^{2}a^{2}LX}+C_{1}\bigg]^{-1},\nonumber\\
\end{eqnarray}
where
\begin{eqnarray}
C_{1}\equiv -4 r^{2}(r+1)\bigg(2k^{2}r+3a^{2}LXm_{g}^{2}\bigg)^{2}\Bigg[a^{2}\bigg(X m_{g}^{2}\big(2k^{2}Lr^{4}\nonumber\\+a^{2}(GH(r+1)\dot{L}+L(3L(r-1)rXm_{g}^{2}-H C_{2}))\big)\nonumber\\-2a^{2}G H^{2}(r+1)M_{GW}^{2}\bigg)\PlanckMass^{2}\Bigg]^{-1},\nonumber\\
\end{eqnarray}
\begin{eqnarray}
C_{2}\equiv -\frac{3r(r+1)}{H^{2}}\bigg[\frac{4H^{2}}{r}-\frac{2k^{2}r}{3a^{2}(r+1)}-\frac{\rho +P}{\PlanckMass^{2}}-\Lambda\bigg],\nonumber\\
\end{eqnarray}
and $\lambda_{2}$ is 
\begin{eqnarray}
\lambda_{2}^{-1}=\frac{3\PlanckMass^{2}}{k^{4}}+\frac{2\PlanckMass^{2}(r+1)}{k^{2}m_{g}^{2}a^{2}LX}+\frac{4\PlanckMass^{2}r^{2}}{3m_{g}^{2}a^{2}}\times \nonumber\\\Bigg\lbrace L Xa^{2}\bigg(-4H^{2}+(\frac{\rho +P}{\PlanckMass^{2}}+\Lambda +m_{g}^{2}LX)\bigg)\nonumber\\+2a^{2}H\big(2H\frac{M_{GW}^{2}}{m_{g}^{2}}-\dot{L}X\big)\Bigg\rbrace^{-1}.
\end{eqnarray}
It can be possible to obtain the $\lambda_{1}$ and $\lambda_{2}$ in the order of $k^{-2}$ in the limit of $k \rightarrow\infty$. So, we have
\begin{eqnarray}
\lambda_{1}\simeq \frac{2(\rho + P)}{\PlanckMass^{4}c_{s}^{2}H^{2}}+\frac{2\Lambda}{\PlanckMass^{2}c_{s}^{2}H^{2}},
\end{eqnarray}
\begin{eqnarray}
\lambda_{2}\simeq \frac{3m_{g}^{2}a^{4}H}{2\PlanckMass^{2}r^{2}}\bigg[\frac{rLX}{2H}\big(\frac{-4H^{2}}{r}+m_{g}^{2}LX\nonumber\\+\frac{\rho +P}{\PlanckMass^{2}}+\Lambda\big)+2H\frac{M_{GW}^{2}}{m_{g}^{2}}-\dot{L}X\bigg].
\end{eqnarray}
The condition of $\lambda_{1}>0$ is the same as the null-energy condition. Thus, the part of $\lambda_{1}$ is related to the matter and is consistent with the ghost-free condition.\\
The eigenvalue $\lambda_{2}$ in the limit of $k\rightarrow\infty$ is related to the scalar graviton. By evaluating the condition of $\lambda_{2}>0$, we can find out that in the limit of dRGT (i.e. $L, \dot{L}\longrightarrow0$), the $\lambda_{2}$ does not lead to zero, which means that the limits of $k\rightarrow\infty$ and dRGT do not commute simultaneously. In fact, the reason for this behaviour can be sought in the expression of $\lambda_{2}$.\\
Finally, it can be shown that by disregarding the terms which would be zero in the limits, we have
\begin{eqnarray}
\lambda_{2}\simeq \frac{m_{g}^{2}}{\PlanckMass^{2}}\bigg[\frac{2(r+1)}{k^{2}a^{2}LX}+\frac{r^{2}m_{g}^{2}}{3a^{2}H^{2}M_{GW}^{2}}\bigg]^{-1}.
\end{eqnarray}
We can conclude that the first expression in the limit of dRGT and the second expression in the limit of $k\rightarrow \infty$ are essential. Therefore, if the conditions are imposed, we have the conditions $\frac{k^{2}L}{a^{2}H^{2}}<<1$ in the limit of dRGT and $\frac{k^{2}L}{a^{2}H^{2}}>>1$ in the limit of $k\rightarrow\infty$.

In this subsection, we have obtained the quadratic action for scalar perturbations. It should be mentioned that scalar perturbations are induced by energy density inhomogeneities. These perturbations are most essential because they show gravitational instability and may lead to the formation of structure in the Universe \cite{Mukhanov:2005sc}. However, such this phenomenological study is beyond the scope of the current paper, thus we leave it for future study.

\section{Conclusion}\label{Con}

It is important to understand new extensions of massive gravity theories, where
gravitational degrees of freedom propagate in a well-behaved manner.  Also, it
is worth examining ghost-free perturbations around their cosmological
backgrounds. In this paper, by introducing a tachyonic field, we have introduced
a new extension of the nonlinear dRGT massive gravity theory.

Here, we have investigated the cosmology and perturbation analysis of the
tachyonic massive gravity theory. We began by proposing a new action and its
total Lagrangian. We presented the background equations for a FLRW background.
Moreover, the self-accelerating background solutions for tachyonic massive
gravity theory were elaborately discussed. In other words, we have shown a way to explain the late-time acceleration of the Universe within the tachyonic massive gravity.

Furthermore, we have presented the tensor perturbation calculation for analysing
the dispersion relation of GWs for the tachyonic massive gravity theory. We have
demonstrated the propagation of gravitational perturbation in the FLRW cosmology. Eventually, these analyses should be compared with the emerging observations of GWs. In particular, the first ever observation of a merging binary neutron stars, namely the GW170817, has demonstrated the usefulness of GWs in analyzing the dispersion relation of GWs~\cite{LIGOScientific:2017zic,LIGOScientific:2018dkp}.

We considered the potential as a constant
$V(\sigma)=\frac{2\omega}{\PlanckMass^{2}}$ in a special case, and we have
obtained the full set of equations of motion for a FLRW background. We have
found self-accelerating background solutions as well.

Finally, for the special case, we have derived the cosmological perturbations,
which contain tensor, vector, and scalar modes. We have pointed out the analysis
of the dispersion relation of GWs, and have presented the details of vector and
scalar perturbations. In fact, we have demonstrated the conditions in which the theory to be stable and ghost-free.

The cosmology and perturbation analysis in this paper might serve as an
interesting starting point for future theoretical and empirical studies of
cosmology and GW data, for example, in the studies of modified propagation of
GWs in an expanding universe~\cite{Will:1997bb,Mirshekari:2011yq,LIGOScientific:2016lio,Shao:2020shv,Nishizawa:2017nef,Nishizawa:2019rra}. These studies are increasingly important nowadays for the field of GW cosmology as more and more relevant GW events are continuously accumulating, and exploring the scientific cases of them rewards the great efforts in making these experiments feasible. From a theoretical viewpoint, it is also worth
mentioning that, following the steps in this work, one could possibly study
other extensions of the nonlinear theory of massive gravity.

\section*{Acknowledgements}
This work has been supported by University of Tabriz, International and Academic Cooperation Directorate, in the framework of TabrizU-300 program. We are really grateful to Nishant Agarwal for helpful notes and codes which are related to tensor perturbations.
Also, A.R.A and S.K would like to thank A.\ Emir Gumrukcuolu for his useful comments. A.R.A would like to thank Y.\ Izadi too.
L.S was supported by the National Natural Science Foundation of China (Grants No. 11975027, No 11991053, No 11721303), the Young Elite Scientists Sponsorship Program by the China Association for Science and Technology (No. 2018QNRC001), the National SKA Program of China (No. 2020SKA0120300), and the Max Planck Partner Group Program funded by the Max Planck Society.


\bibliography{apssamp}


\end{document}